\documentstyle[aps,amsfonts,epsfig]{revtex}
\newcommand{\I}{\text{i}}
\newcommand{\E}{\text{e}}
\newcommand{\tr}{\text{tr}}
\newcommand{\Tr}{\text{Tr}}
\newcommand{\re}[1]{(\ref{#1})}
\newcommand{\sta}[1]{{}^\star\! #1}
\begin{document}
\draft
\twocolumn{
\title{QED effective action at finite temperature}
\author{Holger Gies\thanks{Email address:
    holger.gies@uni-tuebingen.de}} 

\address{Institut f\"ur theoretische Physik\\
          Universit\"at T\"ubingen\\
      Auf der Morgenstelle 14, 72076 T\"ubingen, Germany}
\date{}
\maketitle
\begin{abstract}
  The QED effective Lagrangian in the presence of an arbitrary
  constant electromagnetic background field at finite temperature is
  derived in the imaginary-time formalism to one-loop order. The
  boundary conditions in imaginary time reduce the set of gauge
  transformations of the background field, which allows for a further
  gauge invariant and puts restrictions on the choice of gauge. The
  additional invariant enters the effective action by a topological
  mechanism and can be identified with a chemical potential; it is
  furthermore related to Debye screening. In concordance with the
  real-time formalism, we do not find a thermal correction to
  Schwinger's pair-production formula. The calculation is performed on
% covariant
  a maximally Lorentz covariant and gauge invariant stage.
\end{abstract}
\section{Introduction}

The construction of an effective action for quantum electrodynamics
(QED) in the presence of various external conditions has been a
challenge since the early days of the theory. The study of
generalizations of the Heisenberg-Euler Lagrangian that include
finite-temperature effects has been initiated by Dittrich
\cite{ditt79}, who considered the case of a constant external magnetic
field at finite temperature using the imaginary time formalism. An
extension of this work to the case of arbitrary constant
electromagnetic fields turned out to be qualitatively more substantial
than naively expected. Employing the real-time formalism, this
situation was investigated by Cox, Hellman and Yildiz \cite{cox84},
and Loewe and Rojas \cite{loew92}. A more comprehensive study of the
problem has been performed by Elmfors and Skagerstam \cite{elmf94},
who corrected the preceding findings and additionally introduced a
chemical potential. An attempt employing the imaginary-time formalism
was made by Ganguly, Kaw and Parikh \cite{gang95} for the case of an
external electric field. Recently, the finite-temperature effective
action for electromagnetic fields was studied by Shovkovy
\cite{shov98} in the worldline approach, where finite temperature is
also introduced via an imaginary-time formalism.

This paper is devoted to the derivation of the effective action of
arbitrary constant electromagnetic fields at finite temperature in the
imaginary-time formalism. Similarly to the above-mentioned papers, our
approach is based on Schwinger's proper-time formalism \cite{schw51}
and refers to the one-loop level. By assigning a 4-velocity vector to
%%%%%%%%%%%%%%%%%%%%%%%%%%%%%%%%%%%%%%%%%%%%%%%%%%%%%%
% heat-bath --> heat bath
%%%%%%%%%%%%%%%%%%%%%%%%%%%%%%%%%%%%%%%%%%%%%%%%%%%%%%
the motion of the observer with respect to the heat bath, a manifest
covariant notation is obtained \cite{weld82} which enables us to
formulate the problem in terms of gauge invariant and covariant
quantities.

However, gauge transformations of the finite-tem\-pera\-ture
generating functional are a priori restricted to periodic gauge
functions $\Lambda_{\text{p}}$ in order to leave the boundary
conditions of the functional integral over the fluctuating field
invariant. This requires a more careful choice of gauge for the
background field than at zero temperature, since the reduced class of
gauge transformations allows for more physical information to be
carried in the explicit form of the gauge potential. The additional
information can be associated with a chemical potential.

Apart from subtleties with the correct choice of gauge, we largely
agree with the findings of the ``real-time'' investigations
\cite{elmf94}. We finally comment on the apparent controversy in the
literature concerning the (non-)\-vanishing of the imaginary part of the
thermal effective action that is related to pair production
\cite{cox84}-\cite{hall94}. In concordance with the findings of the
real-time calculations, we do not find a thermal contribution to the
pair-production rate to this order of calculation.

It is, of course, obligatory to point out that the implications of the
present calculation may not be immediately interpretable, since the
presence of an electric field violates the thermal equilibrium
assumption of the imaginary-time formalism. In particular, a constant
electric field transfers energy to thermally fluctuating charged
particles. On a formal level, it is not clear whether the periodicity
in imaginary time can be identified with the physical temperature of
the heat bath. However, there are field configurations allowing for
thermal equilibrium, e.g., a shallow potential well as suggested in
\cite{elmf94}, for which the constant field approximation can be
applicable.

Moreover, the knowledge of the effective action given below depending
on the complete set of invariants of an electromagnetic field
including an additional Lorentz vector (temperature times heat-bath
velocity), might be useful even in the limit of vanishing electric
fields.

\section{Imaginary-Time Formalism}
The one-loop effective action of QED is characterized by the fact that
the fluctuating charged fermions which couple to the external field to
all orders have been integrated out. In this way, finite temperature
is introduced via the imaginary-time formalism by postulating
anti-periodic boundary conditions for these fluctuating fermions in
the direction of imaginary time with period $\beta=\frac{1}{T}$. 

% $2\pi/e$
Regarding the complete generating functional of QED, the external
field is treated as a background field \cite{abbo81}. To maintain
invariance of the fermionic integral under gauge transformations of
the background field, it is important to restrict the gauge functions
$\Lambda(x)$ to be $\beta$-periodic in imaginary time\footnote{In
  principle, one could additionally allow for an integer multiple of
  $2\pi/e$ on the right-hand side of Eq. \re{1.01}. But, since such a
  term does not contribute to the present situation, we will simply
  omit it in the following.},
\begin{equation}
\bigl\{\Lambda_{\text{p}}\bigr\}:\qquad\Lambda_{\text{p}}(x^\mu+\I \beta
u^\mu)=\Lambda_{\text{p}}(x^\mu), \label{1.01}
\end{equation}
where $u^\mu$ denotes the 4-velocity vector of the heat bath. Although
the QED action as well as the integration measure are invariant under
arbitrary gauge transformations $\Lambda(x)$ of the background field,
the anti-periodic boundary conditions will be modified if
$\Lambda(x)\not\in \bigl\{ \Lambda_{\text{p}} \bigr\}$; in particular
$\psi(0)=-\psi(\beta) \to \psi(0)=-\E^{\I e(\Lambda(\beta)-\Lambda(0))}
\psi(\beta)$. At zero temperature, the fermion determinant can only
depend on the field strength $F^{\mu\nu}$ that arises from the
background field; the explicit form of $A_\mu$ is subject to arbitrary
gauge transformations. In contrast, the restricted class of gauge
transformations $\Lambda_{\text{p}}$ at finite temperature allows for
further gauge invariant quantities of the type
\begin{equation}
\bar{A}_u({\mathbf{x}})=\frac{1}{\beta} \int\limits_0^\beta d\tau\,
A_u(x^\mu+\I\tau u^\mu), \qquad A_u:=A^\mu u_\mu, \label{1.02}
\end{equation}
where $\mathbf{x}$ denotes the components of $x^\mu$ orthogonal to
$u^\mu$. Already at this stage, one might suspect that the physical
meaning of $\bar{A}_u$ is related to a chemical potential $\mu$ which
would enter the QED action by adding $\mu \gamma^\mu u_\mu$ to the Dirac
operator: $\Pi\!\!\!\!/=(-\I \partial\!\!\!/- eA\!\!\!/) \to(-\I
\partial\!\!\!/- eA\!\!\!/+ \mu u\!\!\!/)$. In the following, we will
further establish this relation between $\bar{A}_u$ and $\mu$ and
especially demonstrate that the appearance of $\bar{A}_u$ in the
effective action is of topological origin. Instead of employing the
functional integral formalism, we will closely follow Schwinger's
proper-time formalism, which provides for a detailed study of gauge
invariance.

We
therefore begin with the fermionic Green's function in an external
electromagnetic field at zero temperature satisfying the differential
equation
\begin{equation}
[(\gamma^\mu \Pi_\mu) +m]\, G(x,x'|A)= \delta(x-x'), \label{1.1}
\end{equation}
with $\Pi_\mu=-\I\partial_\mu -eA_\mu$. Following Schwinger
\cite{schw51}, we can solve Eq.\re{1.1} formally on an operator level
($G(x,x'|A)=\langle x|G[A]|x'\rangle$):
\begin{equation}
G[A]=(m-\gamma\Pi) \, \I\int\limits_0^\infty ds\,
\E^{-\I m^2s}\, \E^{\I(\gamma\Pi)^2s}. \label{1.2}
\end{equation}
Convergence of this proper-time integral and the following is ensured
by the implicit prescription $m^2\to m^2-\I \epsilon$. The
proper-time transition amplitude
\begin{equation}
K(x,x';s|A):=\langle x|\, \E^{\I s(\gamma\Pi)^2}\, |x'\rangle 
\label{1.3} 
\end{equation}
in the integrand of Eq.\re{1.2} also enters the proper-time formula
for the effective (unrenormalized) one-loop Lagrangian:
\begin{equation}
{\cal L}^{1}=\lim_{x'\to x}\,\frac{\I}{2} \tr_\gamma\,
\int\limits_0^\infty \frac{ds}{s}\, \E^{-\I s m^2} \langle x|\, \E^{\I
  s(\gamma\Pi)^2}\, |x'\rangle .\label{1.4}
\end{equation}
Introducing the scalar propagator
\begin{equation}
\Delta(x,x'|A)=\I \int\limits_0^\infty ds\, \E^{-\I m^2s}\,
K(x,x';s|A), \label{1.5}
\end{equation}
which is related to the fermion's Green's function via
$G[A]=(m-\gamma\Pi) \, \Delta[A]$, we implicitly find an equation for
$K(x,x';s|A)$ which is the Green's function equation for
$\Delta(x,x'|A)$:
\begin{equation}
D[A]\, \Delta(x,x'|A):=[m^2-(\gamma\Pi)^2]\,
\Delta(x,x'|A)=\delta(x-x'), \label{1.6}
\end{equation}
where $D[A]$ abbreviates the differential operator. Obviously,
$K(x,x';s|A)$ as well as the Green's functions $G(x,x'|A)$ and
$\Delta(x,x'|A)$ are gauge dependent. For constant electromagnetic
fields, the solution for the transition amplitude $K(x,x';s|A)$ can
most conveniently be found in the Schwinger-Fock gauge that eliminates
the gauge potential in favor of the field strength:
\begin{equation}
A^\mu_{\text{SF}}:=-\frac{1}{2} F^{\mu\nu}(x-x')_\nu. \label{1.7}
\end{equation}
The solution reads \cite{ditt97}
\begin{equation}
K(x,x';s|A_{\text{SF}})=\!\!\!\int\!\!\!\frac{d^4p}{(2\pi)^4} \E^{-\I
  p(x-x')} \E^{\I \frac{e}{2} \sigma F s} \E^{-Y(\I s)}
\E^{-p{\mathsf{X}}(\I s)p}, \label{1.8}
\end{equation}
where $\sigma F:=\sigma_{\mu\nu}F^{\mu\nu}$,
$\sigma_{\mu\nu}=\frac{\I}{2}[\gamma_\mu, \gamma_\nu]$, and the
quantities $Y$ and $\mathsf{X}$ additionally depend on the field
strength,
\begin{equation}
Y(s)= \frac{1}{2} \text{tr}\, \ln [ \cos (e{\mathsf{F}}s)],\qquad
{\mathsf{X}}(s)= \frac{\tan (e{\mathsf{F}}s)}{e{\mathsf{F}}}, 
\label{1.9} 
\end{equation}
and we used matrix notation, e.g., $F_\mu{}^\nu\equiv
({\mathsf{F}})_\mu{}^\nu$. By insertion of Eq.\re{1.8} into
Eqs.\re{1.5}, \re{1.1} and \re{1.4}, we obtain the explicit
representation for the scalar propagator, the fermion's Green's
function and the effective Lagrangian, respectively, for constant
external fields at zero temperature (in the Schwinger-Fock gauge!).

To introduce finite temperature via the imaginary-time formalism, one
is tempted to replace the $p_0$-integration in Eq.\re{1.8} by a sum
over Matsubara frequencies\footnote{For theories without gauge
  symmetries, of course, this procedure has been applied successfully
  in \cite{grib90}.}. However, this would lead to an incorrect,
or at least incomplete result, since the gauge dependence of the
Green's functions has to be taken into account.

As can be shown, the complete gauge dependence can be treated
multiplicatively by a {\em holonomy} factor. In particular, the
transition amplitude in an arbitrary gauge is related to the one in
the Schwinger-Fock gauge by
\begin{equation}
K(x,x';s|A)=\Phi(x,x'|A)\, K(x,x';s|A_{\text{SF}}), \label{1.10}
\end{equation}
where the holonomy factor reads:
%%%%%%%%%%%%%%%%%%%%%%%%%%%%%%%%%%%%%%%%%%%%
% \int\limits_{x'}^x \xi_\mu\, ---> \int\limits_{x'}^x d\xi_\mu\,
%%%%%%%%%%%%%%%%%%%%%%%%%%%%%%%%%%%%%%%%%%%%
\begin{equation}
\Phi(x,x'|A)=\exp \left[ \I e\int\limits_{x'}^x \!d\xi_\mu\, \left(
    A^\mu(\xi) +\frac{1}{2} F^{\mu\nu}(\xi-x')_\nu\right)
    \right]. \label{1.11}
\end{equation}
Identical relations hold for the Green's functions. Note that the
integrand is curl-free and hence the integral in Eq.\re{1.11} is
path-independent as long as the configuration space is simply
connected. Concerning the effective Lagrangian \re{1.4} at zero
temperature, the holonomy factor plays no role, since $\Phi(x,x'|A)\to
1$ in the coincidence limit $x\to x'$. Consequently, the effective
action is gauge invariant.

The situation changes substantially at finite temperature: since the
imaginary time becomes compactified according to the anti-periodic
boundary conditions, the configuration space is no longer simply
connected. As a consequence, the holonomy factor is only invariant
under continuous deformations of the integration path but can pick up
a winding number by closing the path via the anti-periodic boundary. 

%%%%%%%%%%%%%%%%%%%%%%%%%%%%%%%%%%%%%%%%%%
% remainder of section slightly rewritten 
% correct separation of \Phi(x,x'_n|A) established
% no consequences for the following sections nor conclusions
%%%%%%%%%%%%%%%%%%%%%%%%%%%%%%%%%%%%%%%%%%
The simplest way to establish anti-periodicity in imaginary time is to
apply the method of image sources to the Green's function equation.
Therefore, let $x$ and $x'$ belong to the same topological sector,
i.e., there is a straight path from $x$ to $x'$ which does not cross
the imaginary-time boundaries.  Then we define the reflection points
of $x'$ along the imaginary-time axis by (Fig.1)
\begin{equation}
x'_n=x'-\I\beta nu. \label{1.11a}
\end{equation}

\begin{figure}
\begin{center}
\epsfig{figure=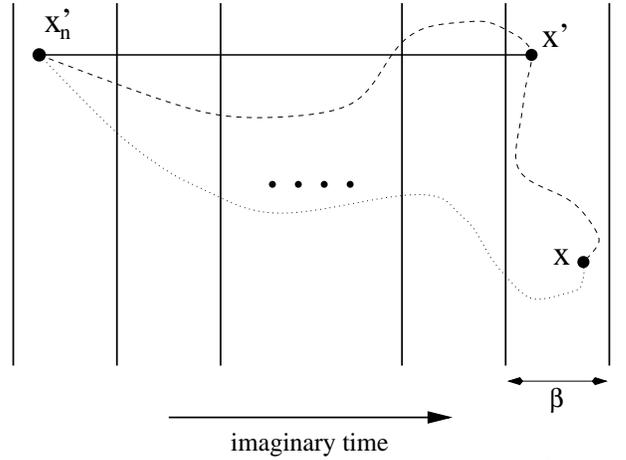,width=8cm}
\caption{The positions of the different points $x$, $x'$ and $x'_n$
  are exhibited. The dotted line represents an arbitrary path from
  $x'_n=x'-\I\beta nu$ to $x$. As a first step, this path is
  continuously deformed in such a way that $x'$ becomes an element of
  the path (dashed line). Secondly, the path from $x'_n$ to $x'$ can
  be deformed to a straight line (solid line), which gives rise to Eq.
  \re{1.15}.}
\end{center}
\end{figure}

Applying the image-source construction, e.g., to Eq. \re{1.6}, we obtain
\begin{eqnarray}
\sum_{n=-\infty}^\infty (-1)^n \delta(x,x'_n)
&&=\sum_{n=-\infty}^\infty (-1)^n\, D[A]\,\Delta(x,x'_n|A) \nonumber\\
&&=D[A]\,\Delta^T(x,x'|A),\label{1.12}
\end{eqnarray}
where $u^\mu$ denotes the 4-velocity of the heat bath, the periodicity
scale is set by the inverse temperature $\beta$, the factor $(-1)^n$
stems from the {\em anti}-periodic boundary conditions, and we have
defined the thermal Green's function
\begin{equation}
\Delta^T(x,x'|A)=\sum_{n=-\infty}^\infty (-1)^n\,
\Delta(x,x'_n|A). \label{1.13} 
\end{equation}
Transition to Fourier space and separation of the
tem\-pera\-ture-de\-pend\-ent parts leads us to
\begin{eqnarray}
\Delta^T(x&&,x'|A)=\!\int\!\! \frac{d^4p}{(2\pi)^4} \E^{-\I p(x-x')}
  \Delta(p)\, \Phi(x,x'|A) \label{1.14}\\
&&\!\times \!\!\sum_{n=-\infty}^\infty\! (-1)^n\, \E^{-\I p(\I\beta
  nu)} \Phi(x',x'_n|A)\, \E^{-\I n\frac{e}{2}\beta (\I
  u{\mathsf{F}}(x-x'))}. \nonumber  
\end{eqnarray}
The separation
\begin{equation}
\Phi(x,x'_n|A)=\Phi(x,x'|A)\,\Phi(x',x'_n|A)\,\E^{-\I
 n\frac{e}{2}\beta (\I u{\mathsf{F}}(x-x'))}\label{1.14a}
\end{equation}
was achieved by a continuous deformation of the integration path in
such a way that, on the one hand, $x'$ becomes an element of the path
and, on the other hand, the path of $\Phi(x,x'|A)$ lies entirely in
the topological trivial sector.

Concerning $\Phi(x',x'_n|A)$, we can deform the integration
path to a straight line (SL) along the imaginary $u^\mu$-direction:
\begin{equation}
\Phi(x',x'_n|A)=\exp \left[\I
  e\int\limits_{\stackrel{{ x'-\I\beta nu}}{\text{SL}}}^{x'}
  d\xi_\mu\,  A^\mu(\xi)\right]. \label{1.15}
\end{equation}

As mentioned above, the exponent in Eq. \re{1.15} is an invariant
quantity under periodic gauge transformations $\Lambda_{\text{p}}$ but
will depend on the explicit form of $A^\mu$ in a certain manner. At
this stage, it is important to point out that the background field
potential is not necessarily subject to periodic boundary conditions,
since it does not correspond to thermalized particles; it is not an
integration variable even in the complete theory. To specify the form
of $A^\mu$, more physical input is required: 
in the present paper, we assume that the system under consideration is
homogeneous. 
Since the effective
Lagrangian for a homogeneous system such as the constant field
configuration has to be independent of $x$, the coincidence limit
$x\to x'$ of the thermal transition amplitude $K^T(x,x';s|A)$ must
also be independent of $x$; the same requirement holds for
$\Delta^T(x,x'|A)$. With regard to Eq. \re{1.14}, this is only
satisfied if $\Phi(x',x'_n|A)$ is independent of $x'$. Thereby, we
obtain the gauge condition
\begin{equation}
0=\int\limits_0^1 d\tau\, \partial_{x'}^\mu A_u(x'{}^\mu-\I\beta
nu^\mu +\tau(\I\beta nu^\mu)). \label{1.15a}
\end{equation}
Condition \re{1.15a} is satisfied if
\begin{equation}
A_u\equiv \bar{A}_u =\text{const.}, \label{1.15b}
\end{equation}
which is the generic choice. Any other solution is gauge equivalent to
Eq. \re{1.15b}, $A_u\to A_u+\partial_u \Lambda_{\text{p}}$. Equation
\re{1.15a} also fixes the choice for the spatial components
$\mathbf{A}$: since $A^\mu$ should produce a constant electric field
via
\begin{displaymath}
\text{const.}={\mathbf{E}}={\mathbf{\nabla}} A_u -\partial_u
{\mathbf{A}} \stackrel{\re{1.15b}}{=} -\partial_u {\mathbf{A}},
\end{displaymath}
the generic choice for $\mathbf{A}$ in the heat bath rest frame
($\partial_u\hat{=} \frac{\partial}{\partial t}$) reads
\begin{equation}
{\mathbf{A}}=-{\mathbf{E}} t+{\mathbf{a}}({\mathbf{x}}), \label{1.15c}
\end{equation}
whereby the function ${\mathbf{a}}({\mathbf{x}})$ is defined by
${\mathbf{B=:\nabla\times a}}$. Again, other choices for
${\mathbf{A}}$ are given by its gauge transforms with respect to
$\Lambda_{\text{p}}$. Note that these gauge conditions are different
from those found in Ref. \cite{elmf94} employing the real-time
formalism.\footnote{In the real-time formalism, the $A^0$-component of
  the gauge field enters the propagators as well as the effective
  action via the thermal distribution function that is given as an
  external condition. Hence, there is no intrinsic criterion for an
  appropriate choice of $A^\mu$, and one has to rely on other
  arguments. E.g., in Ref. \cite{elmf94}, it was argued 
  that a gauge condition of the form $\frac{d}{dt}A^\mu=0$ is required
  for obtaining a clear separation of fermionic and electromagnetic
  energies. This implies that the constant electric background field
  is produced by a spatially non-constant $A^0$,
  ${\mathbf{E=-\nabla}}A^0$, and has to be interpreted
  as a spatially non-constant chemical potential (cf. later). This
  demonstrates that different gauge choices which are not
  gauge-equivalent with respect to $\{\Lambda_{\text{p}}\}$ correspond
  to different physical settings.}
%  the gauge condition $\frac{d}{dt}A^\mu=0$ was erroneously deduced
%  from the conservation of the fermionic energy. However, imposing
%  $\frac{d}{dt}A^\mu=0$ would imply that the constant electric
%  background field must be produced by a spatially non-constant $A^0$:
%  ${\mathbf{E=-\nabla}}A^0$; this, in turn, would lead to a spatially
%  non-constant effective Lagrangian, and would have to be interpreted
%  as a spatially non-constant chemical potential (cf. later).
%  Nevertheless, the important results of Ref. \cite{elmf94}
%  fortunately remain untouched by the problem of gauge choice.}

Taking these considerations into account, the holonomy factor
\re{1.15} eventually yields
\begin{equation}
\Phi(x',x'_n|A)=\exp\bigl[ \I e(\I\beta n)\bar{A}_u
\bigr]. \label{1.18} 
\end{equation}
With the aid of a Poisson resummation, we obtain for the sum in
Eq.\re{1.14}:
%%%%%%%%%%%%%%%%%%%%%%%%%%%%%%%%%%%%%%%%%%%%%%%
% \delta\bigl(p_u&&-e(A-A_{\text{SF}})
%       ----->
% \delta\bigl(p_u&&-e(\bar{A}-A_{\text{SF}})
% 
% 4.2.99
%%%%%%%%%%%%%%%%%%%%%%%%%%%%%%%%%%%%%%%%%%%%%%%
\begin{eqnarray}
\sum_{n=-\infty}^\infty (-1)^n\, \E^{-\I p(\I\beta nu)}
&&\Phi(x',x'_n|A)\,\E^{-\I n\frac{e}{2}\beta (\I u{\mathsf{F}}(x-x'))}
\label{1.19}\\ 
=2\pi\I T \sum_{n=-\infty}^\infty
\delta\bigl(p_u&&-e(\bar{A}-A_{\text{SF}})_u +{\I\pi
  T}(2n+1)\bigr),\nonumber 
\end{eqnarray}
with $p_u=u^\mu p_\mu$, and $A_{\text{SF}\,u}=-\frac{1}{2}u_\mu
F^{\mu\nu}(x-x')_\nu$. Inserting Eq.\re{1.19} into \re{1.14} leads us
to the final expression for $\Delta^T(x,x'|A)$. Similarly, the thermal
transition amplitude $K^T(x,x';s|A)$ as well as the thermal fermion's
Green's function can be derived. Note that these objects contain
temperature-dependent contributions as well as the zero-temperature
part. 
%%%%%%%%%%%%%%%%%%%%%%%%%%%%%%%%%%
% 19.04.99
% thanx to the referee: additional reference
%%%%%%%%%%%%%%%%%%%%%%%%%%%%%%%%%%
The question of gauge dependence of the thermal fermionic Green's
function in a purely magnetic background has also been addressed in
\cite{elmf96}.  

We observe that the Matsubara prescription finally reads
%%%%%%%%%%%%%%%%%%%%%%%%%%%%%%%%%%%%%%%%%%%%%%%
% \case{e(A-A_{\text{SF}})
%       ----->
% \case{e(\bar{A}-A_{\text{SF}})
% 
% 4.2.99
%%%%%%%%%%%%%%%%%%%%%%%%%%%%%%%%%%%%%%%%%%%%%%%
\begin{equation}
\int\!\!\frac{dp_u}{2\pi} f(p_u^2)\to \I T\!\!\!\!\!
\sum_{n=-\infty}^\infty\!\!\! f\Bigl(-\bigl(\pi T(2n+1+{\case{\I}{\pi}  
\case{e(\bar{A}-A_{\text{SF}})_u}{T}})\bigr)^2\Bigr). \label{1.20}
\end{equation}
The explicit appearance of $A_{\text{SF}\,u}$ hints at the fact that
this modified Matsubara prescription will be applied to an object which
has been calculated in the Schwinger-Fock gauge.
%%%%%%%%%%%%%%%%%%%%%%%%%%%%%%%%%%%%%%%%%%%%%
% gauge field shifted ---> gauge field-shifted
%%%%%%%%%%%%%%%%%%%%%%%%%%%%%%%%%%%%%%%%%%%%%
Equation \re{1.20} finally states that it is a gauge field-shifted
momentum in $u^\mu$-direction, $(p-e(A-A_{\text{SF}}))_u$, which is
replaced by Matsubara frequencies instead of the canonical momentum.
This implies a dependence of the Green's functions and the transition
amplitude on the gauge field invariant $\bar{A}_u$ even in the
coincidence limit $x'\to x$ (note that $A_{\text{SF}}\to 0$ for $x'\to
x$).  As a consequence, the effective Lagrangian will be invariant
under periodic gauge transformations $\Lambda_{\text{p}}$ but not
under arbitrary gauge transformations $\Lambda$. Of course, this was
expected from our initial considerations. The physical role of
%%%%%%%%%%%%%%%%%%%%%%%%%%%%%%%%%%%%%%%%%%%%%%%%%%%%%%%
% at the end of the following section.
%  ---->
% at the end of section IV.
%
% 4.2.1999
%%%%%%%%%%%%%%%%%%%%%%%%%%%%%%%%%%%%%%%%%%%%%%%%%%%%%%%
$\bar{A}_u$ will be elucidated at the end of section IV.

\section{Covariant Formulation}

The imaginary-time formalism has often been criticized because it
exhibits the explicit non-covariant feature of leading to discrete
energies but continuous momenta for the quantized fields. In the
present work, we want to demonstrate that it is nevertheless possible
to establish covariance at any stage of this calculation, since the
above-mentioned disproportion between energy and momentum only appears
in internal propagators, all of which are integrated out. Manifest
covariance is achieved by constructing a reference frame that
completely relies on the covariant and gauge invariant building blocks
of the problem.

These building blocks in the present problem of constant
electromagnetic fields at finite temperature are the
field strength tensors, $F^{\mu\nu}$ and $\sta{F}^{\mu\nu}=\frac{1}{2}
\epsilon^{\mu\nu\alpha\beta} F_{\alpha\beta}$; furthermore, we
encounter the heat-bath vector $n^\mu$ \cite{isra76} that is on the
one hand characterized by the value of its invariant scalar
product\footnote{ We employ the metric {\em g}=diag(-1,1,1,1).},
$n^\mu n_\mu=-T^2$, where $T$ denotes the heat-bath temperature, and
on the other hand related to the heat-bath 4-velocity via the
invariant parameter $T$: $n^\mu= T\, u^\mu$. There are 10 independent
components in $F^{\mu\nu}$, $\sta{F}^{\mu\nu}$, and $n^\mu$. The
number of generators of the Lorentz group is 6; hence we can transform
6 components to zero, since there is no little group that leaves
$F^{\mu\nu}$, $\sta{F}^{\mu\nu}$, and $n^\mu$ invariant\footnote{For
  pure EM fields, the dimension of the little group would be 2, since
  boosts along and rotation around the field direction in a system
  where $\mathbf{E}$- and $\mathbf{B}$-fields are parallel leave the
  fields invariant.}.  Therefore, we are left with four Lorentz and
gauge invariant scalars (or pseudo-scalars). For reasons of
convenience, we choose the following set:
\begin{eqnarray}
a&:=&\left( \sqrt{{\cal F}^2+{\cal G}^2}+{\cal
    F}\right)^{\frac{1}{2}}, \nonumber\\ 
b&:=&\left( \sqrt{{\cal F}^2+{\cal G}^2}-{\cal
    F}\right)^{\frac{1}{2}}, \nonumber\\ 
c&:=&\frac{1}{T^2}\bigl(n_\alpha F^{\alpha\mu}\bigr)\bigl(n_\beta
  F^{\beta}{}_\mu\bigr)\equiv\bigl(u_\alpha F^{\alpha\mu}\bigr)
  \bigl(u_\beta F^{\beta}{}_\mu\bigr), \nonumber\\
T&=&\sqrt{-n^\mu n_\mu}.\label{2.1}
\end{eqnarray}
The {\em secular} invariants $a,b$ are related to the solutions to the
secular equation of $F^{\mu\nu}$; e.g., the standard invariants can be
expressed in terms of $a$ and $b$ according to
\begin{eqnarray}
{\cal F}&=&\frac{1}{4}F^{\mu\nu}F_{\mu\nu}=\frac{1}{2}
\bigl({\mathbf{B}^2-\mathbf{E}^2}\bigr) \equiv \frac{1}{2}
\bigl(a^2-b^2), \nonumber\\
|{\cal G}|&=&\left|\frac{1}{4}\sta{F}^{\mu\nu}F_{\mu\nu}\right|
=|-{\mathbf{E\cdot B}}| \,\,\equiv ab. \label{2.2}
\end{eqnarray}
Without loss of generality, we confine ourselves to the case of ${\cal
  G}>0$ (or ${\mathbf{E\cdot B}}<0$) and drop the absolute value
notation. E.g., in a system where $\mathbf{B}$ is anti-parallel to
$\mathbf{E}$, we find: $a=|\mathbf{B}|$ and $b=|\mathbf{E}|$. Note
%%%%%%%%%%%%%%%%%%%%%%%%%%%%%%%%%%%%%%%%%%%%%%%%%%%%%%
% that $c$ is always non-zero,
%   ----->
% that $c$ is positiv-definite,
%
%  4.2.1999                  NOT YET REPLACED !!!!!!!!!!!!!!!!!!!
%%%%%%%%%%%%%%%%%%%%%%%%%%%%%%%%%%%%%%%%%%%%%%%%%%%%%
that $c$ is positive-definite, since $n^\mu$ is a time-like vector;
e.g., in the rest frame of the heat bath, we find $c={\mathbf{E}}^2$.
It is obvious that any gauge invariant Lorentz scalar appearing in the
problem is expressible in terms of this set of invariants \re{2.1}. In
the following, we are going to introduce a convenient coordinate
system in which even the components of any Lorentz vector or tensor of
the problem can be expressed in terms of these invariants. We define
the {\em vierbein} $e^{A\mu}$ which mediates between the given system
labelled by $\mu,\nu,\dots=0,1,2,3$ and the desired system labelled by
the (Lorentz) indices $A,B,\dots=0,1,2,3$ by:
\begin{eqnarray}
e_0{}^\mu&:=& u^\mu,\nonumber\\
e_1{}^\mu&:=& \frac{u_\alpha F^{\alpha\mu}}{\sqrt{c}}, \nonumber\\
e_2{}^\mu&:=& \frac{1}{\sqrt{d}} \bigl( u^\alpha F_{\alpha\beta}
  F^{\beta\mu} -c\, e_0{}^{\mu} \bigr), \nonumber\\
e_3{}^\mu&:=&\epsilon^{\alpha\beta\gamma\mu}\, e_{0\alpha}\,
e_{1\beta}\, e_{2\gamma}, \label{2.3}
\end{eqnarray}
where the quantity $d$ abbreviates the combination of invariants
\begin{equation}
d:=2{\cal F}c-{\cal G}^2+c^2. \label{2.4}
\end{equation}
The vierbein satisfies the identity
\begin{equation}
e_{A\mu}\, e_B{}^\mu =g_{AB}\equiv\text{diag}(-1,1,1,1), \label{2.5}
\end{equation}
where $g_{AB}\sim g^{AB}$ denotes the metric which raises and lowers
capital indices. By a direct computation, we can transform the field
strength tensors and the heat-bath vector:
\begin{eqnarray}
n^A&:=&g^{AB}e_{B}{}^\mu\,n_\mu=(T,0,0,0), \nonumber\\
F_{AB}&:=&e_{A\mu}F^{\mu\nu}e_{B\nu}
  \!=\!\left(\!\!\!\begin{array}{cccc}
    0      &\sqrt{c}   &      0     &      0    \\
  -\sqrt{c}&      0    &\sqrt{d/c}  &      0    \\
    0      &-\sqrt{d/c}&      0     &-{\cal G}/\sqrt{c}\\
    0      &      0    &{\cal G}/\sqrt{c}  &      0    
   \end{array}\!\!\right)\!, \nonumber\\
\sta{F}_{AB}&:=&e_{A\mu}\sta{F}^{\mu\nu}e_{B\nu}
  \!=\!\left(\!\!\begin{array}{cccc}
     0      &-{\cal G}/\sqrt{c}&      0     &\sqrt{d/c} \\
   {\cal G}/\sqrt{c}&      0    &      0     &      0    \\
     0      &      0    &      0     & -\sqrt{c} \\
 -\sqrt{d/c}&      0    &  \sqrt{c}  &      0    
   \end{array}\!\right).\nonumber\\
&& \label{2.8}
\end{eqnarray}
So indeed, the components of these tensors are completely expressed in
terms of invariants. Hence, any tensor algebraic manipulation
involving the objects from Eq.\re{2.8} can immediately be performed
on the level of gauge and Lorentz invariants.

It is worthwhile to point out at this stage that a duality
transformation of the type $\mathbf{E\to B}$ and $\mathbf{B\to -E}$ does
not only imply an interchange of $a$ and $b$ (and a sign flip for
${\cal G}$) but also demands for $c\to c+2{\cal F}=c+a^2-b^2$. Hence,
it is not sufficient in the finite-temperature case to perform a
calculation for magnetic fields and then draw an analogy for electric
fields by replacing $\mathbf{B}\to -\I\mathbf{E}$ -- in contrast to a
zero-temperature calculation.  

\section{Effective Action}
From Eq.\re{1.4}, we can read off the definition of the effective
Lagrangian at finite temperature:
\begin{equation}
{\cal L}^{1+1T}\!\!\!=\frac{\I}{2} \tr_\gamma\,
\int\limits_0^\infty \frac{ds}{s}\, \E^{-\I s m^2} K^T(s|A),
\label{3.1} 
\end{equation}
where the superscript implies that ${\cal L}^{1+1T}$ consists of the
zero-temperature as well as the finite-temperature one-loop part:
${\cal L}^{1+1T}= {\cal L}^1+{\cal L}^{1T}$. The thermal
transition amplitude in the integrand is simply obtained by applying
the modified Matsubara prescription \re{1.20} to the zero-temperature
transition amplitude, Eqs.\re{1.8}-\re{1.10}, in the coincidence
limit:
%%%%%%%%%%%%%%%%%%%%%%%%%%%%%%%%%%%%%%%%%%%%%%%%%%%%%%%%%%%%
%  K^T(s;A)&=&2\I T\!\!
%    ----->
%  K^T(s;A)&=&\I T\!\!
% 6.2.99
%%%%%%%%%%%%%%%%%%%%%%%%%%%%%%%%%%%%%%%%%%%%%%%%%%%%%%%%%%%%
\begin{eqnarray}
K^T(s;A)&=&\I T\!\!\sum_{n=-\infty}^\infty\int\limits_V\!\!
\frac{d^3p}{(2\pi)^3}  \E^{\I \frac{e}{2} \sigma F s} \E^{-Y(\I s)} 
\E^{-p{\mathsf{X}}(\I s)p}\Bigg|,\nonumber\\
&&\qquad\qquad\qquad\qquad\qquad\qquad\!{}^{p_u=e\bar{A}_u-\I\pi
  T(2n+1)}\nonumber\\
&&\label{3.2}
\end{eqnarray}
where $V$ denotes the 3-space volume orthogonal to the
$u^\mu$-direction. We now perform the computation of Eqs.\re{3.1} and
\re{3.2} within the system that we established in the previous
section. With respect to the capital labels, this volume is related to
the components $A,B,\dots=1,2,3$, whereas the components along the
$u^\mu$-direction correspond to the label $A,B,\dots=0$. The
$\mathsf{X}$-matrix in the exponent of Eq.\re{3.2} can now be written
as
\begin{eqnarray}
X_{AB}&\stackrel{\re{1.9}}{=}& \I\Biggl[ \frac{\tan eas}{ea} \frac{(
  b^2g_{AB} -F^2_{AB})}{a^2+b^2} \nonumber\\
&&\qquad+\frac{\tanh ebs}{eb} \frac{(a^2g_{AB}+F^2_{AB})}{a^2+b^2}
  \Biggr], \label{3.3}
\end{eqnarray}
where $F^2_{AB}=F_A{}^CF_{CB}$. Incidentally, the identical equation
also holds, of course, with the labels $A,B$ replaced by $\mu,\nu$,
but then the components are not related to gauge and Lorentz
invariants. The only non-vanishing components of the symmetric tensor
$X_{AB}$ are the diagonal elements as well as $X_{02}$ and
$X_{13}$. The Gaussian momentum integration in Eq.\re{3.2} therefore
results in
\begin{eqnarray}
\int\limits_V\!\!\frac{d^3p}{(2\pi)^3}&&\,\E^{-p^AX_{AB}p^B}
  \label{3.4}\\ 
=&&\frac{\E^{(X_{11}X_{33}-X_{13}^2)\frac{p_0^2}{X_{22}}}}{(4
  \pi)^{\frac{3}{2}}}
  \Bigl(\bigl(X_{11}X_{33}-X_{13}^2\bigr)X_{22}\Bigr)^{-\frac{1}{2}},
  \nonumber
\end{eqnarray}
where we made use of the fact that
$-(X_{00}X_{22}-X_{02}^2)=(X_{11}X_{33}-X_{13}^2)$. Substituting the
modified Matsubara frequencies $p_0\equiv p_u=e\bar{A}_u-\I\pi
T(2n+1)$ into the exponent of Eq.\re{3.4}, the summation over $n$ in
Eq.\re{3.2} can be reorganized according to the Poisson formula:
\begin{equation}
\sum_{n=-\infty}^{\infty}\E^{-\sigma(n-z)^2}=\sum_{n=-\infty}^{\infty}
\left(\frac{\pi}{\sigma}\right)^{\frac{1}{2}}\,\E^{-\frac{\pi^2}{\sigma}
  n^2-2\pi\I\,zn}. \label{3.5}
\end{equation}
In this case, we set $z=-\frac{1}{2} -\frac{\I e\bar{A}_u}{2\pi T}$
and $\sigma=\frac{4\pi^2T^2}{X_{22}}
\bigl(X_{11}X_{33}-X_{13}^2\bigr)$.  At this point, we have to mention
that formula \re{3.5} is not valid for Re $\sigma<0$, which would lead
to a divergent behavior of the sum. This will be checked later on.

The Poisson resummation serves the purpose of sep\-ar\-ating the
zero-\-tem\-pera\-ture from the finite-\-tem\-pera\-ture part, since
the complete loop-momentum in\-te\-gra\-tion/sum\-ma\-tion in \re{3.2}
now yields
%%%%%%%%%%%%%%%%%%%%%%%%%%%%%%%%%%%%%%%%%%%%%%%%%%%%%%%%%%
%   \Bigl(\bigl(X_{11}X_{33}-X_{13}^2\bigr)X_{22}\Bigr)
%  -------->
%     \Bigl(X_{11}X_{33}-X_{13}^2\Bigr)
%
%  6.2.99
%%%%%%%%%%%%%%%%%%%%%%%%%%%%%%%%%%%%%%%%%%%%%%%%%%%%%%%%%%
\begin{eqnarray}
\I T&&\sum_{n=-\infty}^\infty \int\limits_V\!\!\frac{d^3p}{(2\pi)^3}\,
  \E^{-p{\mathsf{X}}p}\label{3.6}\\
&&= \frac{\I}{16\pi^2}
  \Bigl(X_{11}X_{33}-X_{13}^2\Bigr)^{-1} \nonumber\\
&&\quad\times \left[1+2\sum_{n=1}^\infty (-1)^n\E^{-\frac{X_{22}}{
      X_{11}X_{33}-X_{13}^2}\frac{n^2}{4T^2}}\, \cosh
  \frac{e\bar{A}_un}{T}\right]. \nonumber
\end{eqnarray}
Keeping only the ``1'' in the last line leads to the standard
proper-time expression for the zero-temperature effective action,
while the sum represents the thermal correction. Employing the
standard results for the remaining terms in Eqs.\re{3.1} and
\re{3.2},
\begin{eqnarray}
\E^{-Y(is)}&\stackrel{\re{1.9}}{=}&\bigl(\cos eas\, \cosh ebs
  \bigr)^{-1}, \nonumber\\
\tr_\gamma\, \E^{\frac{\I e}{2} \sigma F s}&=&4\cos eas\, \cosh ebs,
  \nonumber
\end{eqnarray}
and inserting the explicit representations of the $X_{AB}$ into Eq.
\re{3.6}, we end up with the desired expression for the one-loop
contribution to the effective Lagrangian for constant electromagnetic
fields at finite temperature:
\begin{eqnarray}
{\cal L}^{1+1T}\!\!\!\!&=&{\cal L}^1+{\cal L}^{1T}, \nonumber\\
{\cal L}^1&=&\frac{1}{8\pi^2} \int\limits_0^\infty \frac{ds}{s^3}
  \,\E^{-\I m^2 s}\,\Biggl[ eas\cot eas \,ebs\coth ebs \label{3.7}\\
&&\qquad\qquad\qquad\qquad-\frac{e^2(a^2-b^2)s^2}{3} -1\Biggr],
  \nonumber\\ 
{\cal L}^{1T}&=&\frac{1}{4\pi^2} \int\limits_0^\infty \frac{ds}{s^3}
  \,\E^{-\I m^2 s}\, eas\cot eas \,ebs\coth ebs \label{3.8}\\
&&\qquad\qquad\times\sum_{n=1}^\infty (-1)^n \E^{\I
  h(s)\frac{n^2}{4T^2}}\, \cosh \frac{e\bar{A}_u n}{T}. \nonumber
\end{eqnarray}
At the zero-temperature part ${\cal L}^1$, we subtracted the divergent
terms, which corresponds to a field strength and charge
renormalization. The function $h(s)$ in the exponent of ${\cal
  L}^{1T}$ is obtained from 
\begin{eqnarray}
h(s)&:=&\frac{\I X_{22}}{X_{11}X_{33}-X_{13}^2} \label{3.9}\\
&=&\frac{b^2-c}{a^2+b^2}\, ea \cot eas +\frac{a^2+c}{a^2+b^2}\,
eb\coth ebs. \nonumber
\end{eqnarray}
In the rest frame of the heat bath where $c={\mathbf{E}}^2$, we
recover the findings of Ref. \cite{elmf94} for $h(s)$. Note that
$h(s)$ is strictly real, so there are no apparent convergence problems
in employing the Poisson resummation \re{3.5}. However, it is not a
straightforward exercise to obtain numerical estimates for Eq.
\re{3.8}, due to the wildly oscillatory behavior of the whole
integrand, especially in the sum. Let us for the moment remark that
$h(s)$ reduces to $\frac{1}{s}$ in the limit of vanishing electric
fields (the limit is most conveniently taken for $\mathbf{E}$ and
$\mathbf{B}$ (anti-)parallel). And assuming $\bar{A}_u=0$, we recover
the findings of Ref. \cite{ditt79} for a purely magnetic field at
finite temperature. Furthermore, the general form of Eq.\re{3.8}
coincides with the representation found in the worldline approach
\cite{shov98} (in the heat-bath rest frame), except for the dependence
on the gauge potential; the importance of the holonomy factor has been
overlooked in \cite{shov98}.

%%%%%%%%%%%%%%%%%%%%%%%%%%%%%%%%%%%%%%%%%%%%%%%
% The physical interpretation can most 
%   ---->
% The physical interpretation of $\bar{A}_u$ can most 
%
%  6.2.99
%%%%%%%%%%%%%%%%%%%%%%%%%%%%%%%%%%%%%%%%%%%%%%%
The physical interpretation of $\bar{A}_u$ can most easily be
illuminated in the limiting case of vanishing field invariants,
$a,b,c=0$; under these circumstances we are able to rotate the contour
$s\to -\I s$ and an interchange of integration and summation is
permitted in Eq.\re{3.8}:
\begin{eqnarray}
{\cal L}^{1T}(\bar{A}_u)&=&-\frac{1}{4\pi^2}\int\limits_0^\infty
  \frac{ds}{s^3} \E^{-m^2s} \sum_{n=1}^{\infty} (-1)^n
  \E^{-\frac{n^2}{4T^2s}} \cosh {\scriptstyle \frac{e\bar{A}_un}{T}}
  \nonumber\\ 
&=&-\frac{2}{\pi^2} m^2T^2\sum_{n=1}^\infty \frac{(-1)^n}{n^2} \cosh
  {\scriptstyle \frac{e\bar{A}_un}{T}}\, K_2\bigl( {\scriptstyle
  \frac{m}{T} n} \bigr)\nonumber\\
&=&-\frac{2T}{\pi^2} \sum_{n=1}^\infty \frac{(-1)^n}{n} \cosh
  {\scriptstyle \frac{e\bar{A}_un}{T}}\, \int\frac{d^3k}{4\pi}
  \E^{-\frac{\sqrt{k^2+m^2}}{T}n} \nonumber\\
&=&T\,2\int \frac{d^3k}{(2\pi)^3}\Biggl[ \ln \biggl( 1+\E^{-\beta
  (E+e\bar{A}_u)}\biggr) \nonumber\\
&&\qquad\qquad\qquad +\ln \biggl( 1+\E^{-\beta
  (E-e\bar{A}_u)}\biggr) \Biggr], \label{3.10}
\end{eqnarray}
where we introduced the particle energy for the $e^+e^-$-gas,
$E:=\sqrt{k^2+m^2}\equiv |k_u|$, and two different representations of
the modified Bessel function $K_2$ were taken advantage of
\cite{GR}. According to the relation 
\begin{displaymath}
{\cal L}^{1T}(\bar{A}_u)=T\, \frac{\ln Z(T,\bar{A}_u)}{V},
\end{displaymath}
we indeed find the general expression for the partition function $Z$
of an ideal $e^+e^-$-gas in which the $\bar{A}_u$-field plays the role
of a chemical potential. If we had started the computation including a
chemical potential, we would always have encountered the combination
$-e\bar{A}_u-\mu\hat{=}e\bar{A}^0-\mu$, which is therefore the only
physical quantity.  In other words, we can identify $e\bar{A}_u$ with
a chemical potential during the complete calculation; hence, the
additional information (compared to the zero temperature case) which
is required to define the correct choice of the background gauge
potential $A^\mu$ is obtained from the value of the chemical potential
of the system under consideration. If one wants to perform a gauge
transformation beyond the class of periodic $\Lambda_{\text{p}}$, one
has to redefine the chemical potential to obtain the same physical
system.

The case of the effective Lagrangian of a constant magnetic field at
non-zero chemical potential has been discussed in \cite{chod90}. Based
on the real-time formalism, a comprehensive study of this situation
including finite temperature can be found in \cite{elmf93} where also
astrophysical implications are discussed. The same physical situation
was investigated employing the imaginary-time formalism in
\cite{cang96} where high- and low-temperature expansions were
approached in a more direct way. As is demonstrated in these
references, the zero-temperature limit of Eq.\re{3.8} at a chemical
potential obeying $\mu>m$ requires careful study. 

%%%%%%%%%%%%%%%%%%%%%%%%%%%%%%%%%%%%%%
% 19.04.99
% Thanx to the referee: additional reference
%%%%%%%%%%%%%%%%%%%%%%%%%%%%%%%%%%%%%%
A detailed weak-field expansion of the effective Lagrangian at 
finite temperature and chemical potential was performed in
\cite{elmf98}, relying on the ``real-time'' representation of the
effective action as given in \cite{elmf94}.

\section{Discussion}
Going beyond the constant field approximation, the effective
Lagrangian in Eqs. \re{3.7} and \re{3.8} can be viewed as the zeroth
order of a gradient expansion of the one-loop effective action which
governs the dynamics of the background gauge field $A^\mu(x)$. An
immediate physical consequence of the fact that the $\bar{A}_u$-field
appears explicitly in the Lagrangian is the well-known Debye screening
%%%%%%%%%%%%%%%%%%%%%%%%%%%%%%%%%%%%%%%%%%%%%%%%%%%%%%%%%%
% global minus sign
%
%  6.2.99
%%%%%%%%%%%%%%%%%%%%%%%%%%%%%%%%%%%%%%%%%%%%%%%%%%%%%%%%%%
of electric fields. A weak-field expansion will take the form ${\cal
  L}^{1T}=-\frac{1}{2} \partial_\mu \bar{A}_u\partial^\mu \bar{A}_u
+\frac{m_{\text{eff}}^2}{2} (\bar{A}_u)^2+ {\cal O}(\bar{A}_u^4)$,
where the effective photon mass (inverse Debye screening length) is
given by
\begin{equation}
m_{\text{eff}}^2(T)=\frac{\partial^2{\cal L}^{1T}}{\partial
  \bar{A}_u^2} \Biggl|_{\bar{A}_u=0}. \label{4.1}
\end{equation}
Considering the zero-field limit for simplicity, we find
\begin{equation}
\frac{\partial^2{\cal L}^{1T}}{\partial \bar{A}_u^2}
\Biggl|_{\bar{A}_u=0}= -\frac{2e^2}{\pi^2}\, m^2\sum_{n=1}^\infty
(-1)^n\, K_2\bigl( {\scriptstyle \frac{m}{T}n}\bigr), \label{4.2}
\end{equation}
where $K_2$ denotes a modified Bessel function. In the
high-temperature limit, $T\gg m$, the sum can be expanded, e.g.,
employing the techniques described in the Appendix B of \cite{ditt98},
leading to $\sum_{n=1}^\infty (-1)^n\, K_2\bigl( {\scriptstyle
  \frac{m}{T}n}\bigr)\simeq \frac{\pi^2T^2}{6m^2}+{\cal O}(1)$. We
finally arrive at
\begin{equation}
m _{\text{eff}}^2(T)=\frac{(eT)^2}{3}, \label{4.3}
\end{equation}
which is the well-known result found in the literature. 
%%%%%%%%%%%%%%%%%%%%%%%%%%%%%%%%%%%%
% 19.04.99
% Thanx to the referee: Additional citation
%%%%%%%%%%%%%%%%%%%%%%%%%%%%%%%%%%%%
The leading corrections to the Debye mass in the high-density and
high-temperature limit can be looked up in the Erratum of Ref.
\cite{elmf94}.

At last we turn to the question of whether Schwinger's famous formula
for the pair-production probability renders finite-temperature
corrections at one-loop order. While no thermal contributions have
been found to the imaginary part of ${\cal L}^{1T}$ in \cite{cox84} or
\cite{elmf94} within the real-time formalism nor in \cite{hall94}
employing the functional Schr\"odinger representation, an imaginary
part seems to appear in the imaginary-time formalism \cite{gang95}.
Besides, the latter result had also been computed in the real-time
formalism in \cite{loew92}.

Although our findings for the effective thermal Lagrangian
Eq.\re{3.8} in the heat-bath rest frame formally coincide with those
found in \cite{loew92} (up to numerical pre-factors and an interchange
of proper-time integration and summation), we do not agree with their
computation of the imaginary part, which follows the line of the
zero-temperature calculation. Various obstacles are encountered when
proceeding in this way for the finite-temperature case: since the
function $h(s)$ in the exponential of Eq.\re{3.8} reduces to 
\begin{equation}
h(s)=eE\, \coth eEs \label{4.4}
\end{equation}
for a purely electric field, $E=|\mathbf{E}|$, (i) a rotation of the
contour, $s\to -\I s$, becomes useless due to the $\coth$ term in the
exponent of Eq.\re{3.8}; (ii) each term in the sum of Eq.\re{3.8}
exhibits an essential singularity at the poles of the $\coth$ term on
the imaginary axis; the use of the rule $\cot z\to \text{P}\cot
z+\I\pi\sum_{z_0}\delta(z- z_0)$ is therefore senseless (cf.
\cite{loew92}); (iii) proper-time integration and summation must not
be interchanged. If they are, the imaginary parts of the successive
terms in the sum diverge exponentially, as can be shown by evaluating
the residues of the singularities on the imaginary $s$-axis.
Incidentally, we do not agree with the imaginary part computed in
\cite{gang95}, simply because the expressions for the effective
Lagrangians do not coincide.

However, we can give an indirect argument for the vanishing of the
imaginary part following Ref.\cite{elmf94}: due to the formal
resemblance between our result Eq.\re{3.8} and the findings of Loewe
and Rojas \cite{loew92} for the effective Lagrangian (not for its
imaginary part!), we can follow their steps backwards and end up with
the starting point of the real-time formalism,
\begin{equation}
\frac{\partial\Gamma^{1T}}{\partial m} =-\I\Tr\, \left\{\!
  f_{\text{F}}(k_u,\bar{A}_u)
  \left(\!\frac{1}{\Pi\!\!\!\!/-m+\I\epsilon}-
  \frac{1}{\Pi\!\!\!\!/-m-\I\epsilon}\!\right)\!\right\}, \label{4.5} 
\end{equation}
where $f_{\text{F}}(k_u,\bar{A}_u)$ denotes a (real) thermal
distribution function for the fermions and $\Gamma=\int d^4x\, {\cal
  L}^{1T}$. Obviously, since the right-hand side is purely real, there
is no imaginary part in the thermal contribution to the effective
one-loop action and hence no thermal correction to the Schwinger
pair-production formula to this order of calculation.

\section{conclusion}
In the present work, we studied the derivation of the effective QED
action to one-loop order in presence of arbitrary constant
electromagnetic fields at finite temperature in the imaginary-time
formalism. Although the final expression for the effective action is
well known and has been studied extensively, especially in the
real-\-time formalism \cite{elmf94}, the problem as treated in the 
imaginary-time formalism reveals some delicate features.

Gauge invariance of the classical action turns out to be restricted to
periodic gauge transformations $\Lambda_{\text{p}}$ on the quantum
level in order to leave the boundary conditions of the functional
integral unaltered. This implies the existence of further gauge
invariant quantities beside the field strength which are constructed
from the background field $A^\mu$. Additional information about the
system under consideration has to be employed to fix the form of the
gauge potential $A^\mu$. In the present case, the demand for
homogeneity (constant fields and constant chemical potential) gives rise
to the additional gauge invariant quantity $\bar{A}_u$. 

The way in which $\bar{A}_u$ enters the effective action can be viewed
as a topological effect that arises from the compactification of the
finite-temperature configuration space in imaginary time; the
configuration space, namely, loses its property of being simply
connected and allows for infinitely many topologically inequivalent
paths to connect two different points in space-time. Each path can be
classified by its winding number around the space-time cylinder. The
holonomy factor that carries the gauge dependence of the Green's
function is sensitive to these inequivalent paths, since it represents
a mapping of the paths in configuration space into the gauge group. A
Poisson resummation of the sum over the winding number leads to a sum
over Matsubara frequencies shifted by the $\bar{A}_u$-field.  The
quantity $e\bar{A}_u$ indeed acts like a chemical potential in the
partition function and therefore can be identified with $\mu$.

The gauge non-invariance of the effective action under non-periodic
gauge transformations furthermore manifests itself in giving a mass to
the (integrated) time-like gauge field component $\bar{A}_u$, which
is, of course, nothing but the screening mass of the Debye mechanism.

As a second focus of this work, we introduced a certain coordinate
frame related to the given Lorenz vectors and tensors of the problem
that allowed for a manifest covariant computation in the sense of
relativistic thermodynamics. This procedure helped us to present the
result in terms of a complete set of Lorenz and gauge invariants.

As an immediate application of the final formula for the effective
Lagrangian, we discussed a possible thermal contribution to the
Schwinger pair-production formula. In agreement with the results of
the real-time formalism, we do not find an imaginary part in the
thermal contribution to the effective Lagrangian to this order of
calculation. On a heuristic level, it appears plausible that a thermal
contribution to pair production can arise from higher loop graphs.
E.g., the two-loop process contains the mass operator (in the presence
of an external field), which can be associated with collective
excitations at finite temperature. These can be approximately taken
into account by replacing the fermion mass by an effective $T$ or
$\mu$ dependent mass \cite{peti92}. However, since these effective
masses generally exceed the fermion mass, such thermal contributions
are expected to be of subdominant importance.

\acknowledgements

The author would like to thank Prof. W. Dittrich for insightful
discussions and for carefully reading the manuscript. Helpful comments
by K. Langfeld, M. Engelhardt, and O. Tennert are also gratefully
acknowledged.

\end{document}